\documentclass[twocolumn,showpacs,prl]{revtex4}
\usepackage{amssymb}

\usepackage{graphics}
\usepackage{epsfig}

\usepackage{dcolumn}
\usepackage{amsmath}
\usepackage{color}

\newcommand{\beq}{\begin{equation}}
\newcommand{\eeq}{\end{equation}}
\newcommand{\bea}{\begin{eqnarray}}
\newcommand{\eea}{\end{eqnarray}}
\newcommand{\bwt}{\begin{widetext}}
\newcommand{\ewt}{\end{widetext}}

\begin{document}

\title{Rectification at Graphene-Semiconductor Interfaces: Zero-Gap Semiconductor Based Diodes}
\date{\today}
\author{S. Tongay$^{1,2,3}$, M. Lemaitre$^{1}$, X. Miao$^{2}$, B. Gila$^{2,3}$, B. R. Appleton$^{2,3}$,
and A. F. Hebard$^{2}$}

\begin{abstract}
Using current-voltage ($I$-$V$), capacitance-voltage ($C$-$V$) and electric field modulated Raman measurements, we report on the unique physics and promising technical applications associated with the formation of Schottky barriers at the interface of a one-atom-thick zero-gap semiconductor (graphene) and conventional semiconductors. When chemical vapor deposited graphene is transferred onto \textit{n}-type Si, GaAs, 4H-SiC and GaN semiconductor substrates, there is a strong van der Waals attraction that is accompanied by charge transfer across the interface and the formation of a rectifying (Schottky) barrier. Thermionic emission theory in conjunction with the Schottky-Mott model within the context of bond-polarization theory provides a surprisingly good description of the electrical properties. Applications, such as to sensors where in forward bias there is exponential sensitivity to changes in the Schottky barrier height due to the presence of absorbates on the graphene or to analogue devices for which Schottky barriers are integral components are promising because of graphene's mechanical stability, its resistance to diffusion, its robustness at high temperatures and its demonstrated capability to embrace multiple functionalities. 
\end{abstract}
\affiliation{
$^1$ Materials Science and Engineering, University of Florida, Gainesville, Florida 32611}
\affiliation{
$^2$ Department of Physics, University of Florida, Gainesville, FL 32611}
\affiliation{
$^3$ Nanoscience Institute for Medical and Engineering Technology, University of Florida, Gainesville, FL 32611}

\maketitle

\section{1. INTRODUCTION}

Single atom layers of carbon (graphene) have been studied intensively after becoming experimentally accessible with techniques such as mechanical exfoliation\cite{novoselov}, thermal decomposition on SiC substrates\cite{berger} and chemical vapor deposition (CVD) \cite{ruoff,Hong}. Graphene is a zero-gap semiconductor (ZGS) with an exotic linearly dispersing electronic structure, high optical transparency, exceptional mechanical stability, resilience to high temperatures and an in-plane conductivity with unusually high mobility\cite{castroneto}.  Accordingly, graphene has been proposed as a novel material for incorporation into devices ranging from Schottky light emitting diodes (LEDs) \cite{Li1,Fan,Cronin} to field effect transistors (FETs) \cite{Yi,Phaedon}. Although integration of graphene into semiconductor devices is appealing, there is still very little known about the interface physics at graphene-semiconductor junctions. To this end, graphene/Si junctions showing successful solar cell operation have been produced by transferring either CVD-prepared~\cite{Li1} or exfoliated~\cite{Cronin} graphene sheets onto Si substrates. The resulting diodes have shown ideality factors (measure of deviation from thermionic emission) varying from $\sim$1.5 ~\cite{Li1} which is close to the ideal value of unity, to values in the range $\sim$5-30 on exfoliated graphene~\cite{Cronin} implying that additional non-thermionic current carrying processes exist at the graphene/Si interface. Nevertheless these promising results point to the need for additional research on integrating graphene with technologically important semiconductors. 

\begin{figure}[b]
\includegraphics[angle=0,width=0.5\textwidth]{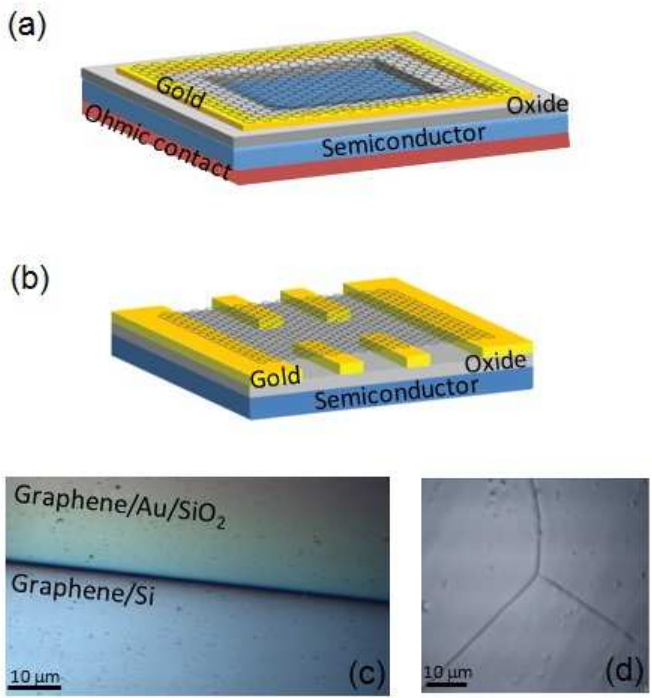}
\caption{{\footnotesize (a) Graphene/semiconductor diode sample geometry where the $J-V$ characteristics were measured between ohmic contact (ground) and graphene (high) (b) Hall bar geometry for measurements of the carrier density of graphene. In this configuration the graphene does not make contact with the semiconductor.} (c) Optical image of the graphene/Au/SiO$_2$ - graphene/Si transition edge after the graphene transfer. (d) Scanning electron microscope image of Cu foils after the CVD graphene growth showing formation of grain sizes large with respect to the 10 $\mu$m scale bar.}
\label{Diagram}
\end{figure}

Here we report rectification (diode) effects at ZGS-semiconductor (i.e graphene-semiconductor) interfaces on a surprisingly wide variety of semiconductors. In addition to current-voltage measurements we utilize, Hall, capacitance-voltage and electric field modulated Raman techniques to gain heretofore unrecognized insights into the unique physics occurring at the graphene/semiconductor interface. We find that when CVD-prepared graphene sheets are transferred onto \textit{n}-type Si, GaAs, 4H-SiC and GaN semiconductor substrates, equilibration of the Fermi level throughout the system gives rise to a charge transfer between the graphene and semiconductor, thereby creating strong rectification (Schottky effect) at the interface. We find that graphene's Fermi level ($E_F^{gr}$) is subject to variation during charge transfer across the graphene-semiconductor interface as measured by \textit{in-situ} Raman spectroscopy measurements, unlike conventional metal-semiconductor diodes where the Fermi level (E$_F$) of the metal stays constant due to a high density of states at the Fermi level. These effects become particularly pronounced at high reverse bias voltages when the induced negative charge in the graphene is sufficient to increase $E_F^{gr}$ and give rise to increased current leakage. 
Our observations and interpretation based on a modification of thermionic emission theory not only provide a new understanding for the development of high frequency, high power, and high temperature Schottky based devices, such as metal-semiconductor field effect transistors (MESFETs) and high electron mobility transistors (HEMTs), but also allow us to integrate graphene into semiconductor devices while simultaneously preserving the superior properties of the graphene and avoiding chemical-structural modifications to the semiconductor.

\section{2. EXPERIMENTAL METHODS}

Our diodes are fabricated by transferring large scale graphene sheets grown by chemical vapor deposition (CVD) directly onto the semiconductor under investigation and allowing Van der Waals attraction to pull the graphene into intimate contact with the semiconductor. Large-area single layer graphene sheets were synthesized on Cu foils via a multi-step, low-vacuum CVD process similar to that used in Ref.~\cite{ruoff2}.  A quartz tube furnace operating in CVD-mode was loaded with 25-50~$\mu$m-thick Cu foils (Puratronic, 99.9999\% Cu), evacuated to 4~mTorr, and subsequently heated to 500$^{\circ}$C under a 25~sccm flow of H$_{2}$ at 325~mTorr. After 30 minutes soak, the temperature was raised to 1025$^\circ$C for 60~minutes to promote Cu grain growth (mean grain size exceeds 5~mm$^2$ determined by optical microscopy). An initial low-density nucleation and slow growth phase was performed at 1015$^{\circ}$C for 100~minutes with a mixture of CH$_{4}$ and H$_2$ at a total pressure of 90~mTorr and flows of $\leq$~0.5 and 2~sccm, respectively.  Full coverage was achieved by dropping the temperature to 1000$^{\circ}$C for 10 minutes while increasing the total pressure and methane flow to 900~mTorr and 30~sccm, respectively.  A 1.5~$\mu$m-thick film of PMMA (MicroChem, 11$\%$ in Anisole) was then spin-cast onto the Cu foils at 2000~rpm for 60 seconds.   The exposed Cu was etched in an O$_2$ plasma to remove unwanted graphene from the backside of the samples.  The PMMA supported films were then etched overnight in a 0.05~mg/L solution of Fe(III)NO$_3$ (Alfa Aesar) to remove the copper. The graphene-PMMA films were then washed in deionized water, isopropyl alcohol (IPA), and buffered oxide etch for 10 minutes, each. After growth and transfer, the graphene films were characterized and identified using a Horiba-Yvon micro-Raman spectrometer with green, red and UV lasers.

Commercially available semiconducting wafers were purchased from different vendors. $n$-Si and $n$-GaAs samples were doped with P (2-6$\times$$10^{15}$~cm$^{-3}$) and Si (3-6$\times$$10^{16}$~cm$^{-3}$) respectively. Epilayers of $n$-GaN and $n$-4H-SiC, 3-6~$\mu$m-thick, were grown on semi-insulating sapphire substrates with Si (1-3$\times$$10^{16}$~cm$^{-3}$) and N (1-3$\times$$10^{17}$~cm$^{-3}$) dopants. During the sample preparation and before the graphene transfer, the wafers were cleaned using typical surface cleaning techniques. Ohmic contacts to the semiconductors were formed using conventional ohmic contact recipes\cite{han,sze,hao,ruvimov}. Multilayer ohmic contacts were thermally grown at the back/front side of the semiconductor and were annealed at high temperatures using rapid thermal annealing. After the ohmic contact formation, a 0.5-1.0~$\mu$m thick SiO$_x$ window was grown on various semiconductors using a plasma enhanced chemical vapor deposition (PECVD) system, and $\sim$~500~nm thick gold electrodes were thermally evaporated onto SiO$_x$ windows at 5$\times 10^{-7}$ Torr. The graphene contacting areas were squares with sides in the range 500\thinspace $\mu$m to 2000\thinspace $\mu$m. Application of IPA improves the success rate of the graphene transfer and does not affect the measurements presented here. After depositing the graphene/PMMA films, the samples were placed in an acetone vapor rich container for periods ranging from 10 minutes to ~$\sim$10 hours. The acetone bath allows slow removal of the PMMA films without noticeable deformation of the graphene sheets. 

Prior to the graphene transfer there is an open circuit resistance between the Au contacts and the semiconductor. After the transfer of the PMMA-graphene bilayer, the graphene makes simultaneous connection to the Au contacts and the semiconductor as evidenced by the measured rectifying I-V characteristics. Since the diodes made with the PMMA-graphene bilayer show essentially the same rectifying characteristics as the samples in which the PMMA has been dissolved away, we conclude that the carbon layer on the PMMA (shown by Raman measurements to be graphene) is making intimate contact with the semiconductor.

\begin{figure}[b]
\includegraphics[angle=0,width=0.5\textwidth]{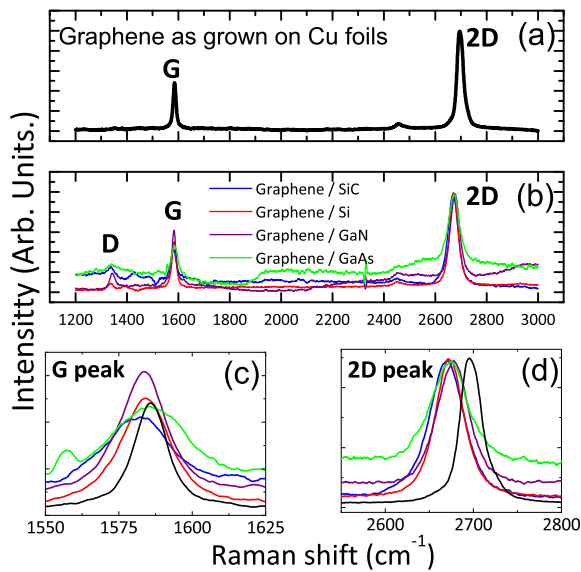}
\caption{{\footnotesize (a) Raman spectra of CVD-grown graphene on Cu foils and (b) graphene after transfer onto various semiconductor substrates. Graphene sheets show large $I_{2D}/I_{G}$ ratio, and after the transfer the graphene becomes slightly disordered. (c-d) Raman G and 2D peaks measured respectively on graphene/Cu and on graphene/semiconductor combinations indicated in the legend of panel (b).}}
\label{Raman}
\end{figure}
											
A schematic for our graphene based diodes is shown in Fig.~\ref{Diagram}(a); the backside of the semiconductor substrate is covered with an ohmic contact and the graphene sheet transfered onto Cr/Au contacts grown on SiO$_{x}$ windows. After the transfer, the graphene and semiconductor adhere to each other in an intimate Van der Waals contact in the middle of the open window, and the Cr/Au contact pad provides good electrical contact with the graphene. Our ohmic contact arrangements allow current density versus voltage ($J$-$V$) and capacitance versus voltage ($C$-$V$) measurements to be taken separately. $J$-$V$ measurements were taken in dark room conditions using a Keithley 6430 sub-femptoamp source-meter, and $C$-$V$ measurements were taken using an HP 4284A capacitance bridge. The electric field modulated Raman measurements were made on the same configuration. Four-terminal transport and Hall measurements however were performed with an intervening layer of SiO$_x$ (Fig.~\ref{Diagram}(b)) using a physical property measurement system (PPMS), at room temperature in magnetic fields up to 7 Tesla.

\section{3. RESULTS AND DISCUSSION}
\subsection{A. Raman measurements}

\begin{figure}[h]
\includegraphics[angle=0,width=0.5\textwidth]{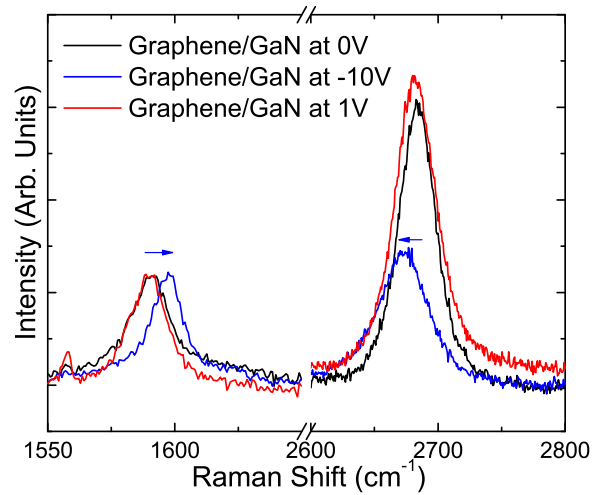}
\caption{{\footnotesize In-situ Raman spectra taken on Graphene/GaN junctions as a function of applied bias: 0V (black line), +1V (red line) and -10V (blue line).}}
\label{RamanBias}
\end{figure}

In Fig.~\ref{Raman}(a-d), we show typical Raman spectroscopy data taken on graphene sheets grown onto Cu foils by CVD deposition before and after transferring onto semiconductors. The presented scans have been reproduced at more than 20 random spots and are good representations of the quality of the graphene on the Cu foils before transfer and on the semiconductor surface after transfer. In the literature the quality of graphene sheets is measured by a large 2D to G intensity ratio ($I_{2D}$/$I_{G}$) and a low D peak intensity ($I_{D}$). Single layer graphene is expected to show $I_{2D}$/$I_{G} > 2$, and the amount of disorder in the sheets is often correlated with $I_{D}$. In our samples, we observe $I_{2D}$/$I_{G}\geq 2$ and a negligible D peak amplitude. However after graphene transfer to the semiconductor substrate, we observe that $I_{D}$ becomes apparent while $I_{2D}$/$I_{G}$ remains the same (Fig.~\ref{Raman}(b)). The increase in $I_{D}$ reflects the lower sheet mobility of CVD-grown graphene and gives rise to weak localization effects at low temperatures~\cite{Cao}. Moreover, because of the low solubility of carbon in Cu, graphene growth onto Cu foils is known to be self-limiting~\cite{ruoff} therefore allowing large-area single layers of graphene to be grown onto Cu foil surfaces. After the graphene growth, the backside of the Cu foils have been exposed to O$_2$ plasma to remove unwanted graphene and checked with Raman spectroscopy. This step assures that bi-layer (or multi-layer) graphene is not formed on PMMA/graphene after etching the Cu foils (see Experimental methods).

The Raman spectrum of exfoliated graphene transferred onto Si/SiO$_2$ substrates has been previously studied as a function of applied bias~\cite{Sood}. It has been found that the G and 2D peaks of graphene are sensitive to the Fermi energy (carrier density) of graphene and allow one to estimate the bias-induced changes in E$_{F}^{gr}$. Considering the typical operating voltages of Schottky junctions, the low carrier density in graphene, and the associated bias dependence of E$_{F}^{gr}$, we have also measured the Raman spectrum of graphene transferred onto GaN as a function of applied bias. Our Raman measurements differ from those reported in Ref.~\cite{Sood} in the following three ways: (1) we  are using CVD-prepared rather than exfoliated graphene, (2) the graphene is in direct contact with GaN rather than oxidized Si, and (3) the graphene is measured \textit{in situ} as part of a Schottky rather than a gated FET. 
In Fig.~\ref{RamanBias}, we show the evolution of the Raman spectrum as a function of applied bias. While G and 2D are almost identical with the same peak positions at 0V and 1V, in reverse bias at 10V, the G band shifts to higher (by $\sim$6$\thinspace$$\pm$3$\thinspace$cm$^{-1}$) and the 2D band shifts to lower (by $\sim$7$\thinspace$$\pm$3$\thinspace$cm$^{-1}$) wavenumbers. The relative shifts in the Raman peaks along with a reduction of the 2D/G peak ratio from 2.6 (at 0$\thinspace$V) to 1.2 (at 10$\thinspace$V) imply that graphene sheets transferred onto GaN become electron doped. Considering the previous results reported on graphene/SiO$_2$~\cite{Sood}, the shift in E$_F$ can be estimated to be in the range $\sim$0.2-0.5$\thinspace$eV. 

\subsection{B. Hall measurements}
\begin{figure}[b]
\includegraphics[angle=0,width=0.5\textwidth]{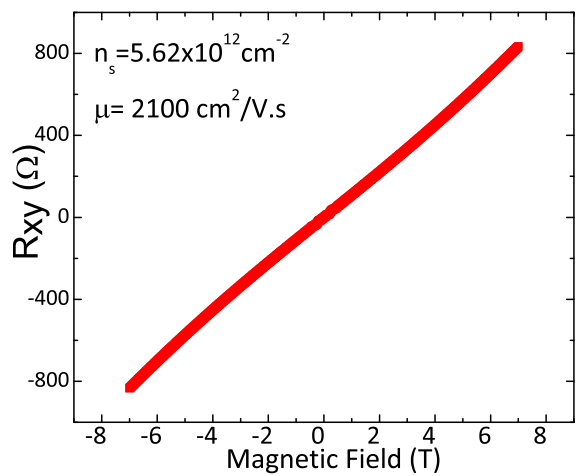}
\caption{{\footnotesize R$_{xy}$ versus magnetic field data taken at 300~K. Typically sample mobilities are in the range 1400-2100 cm$^{2}$/Vs and carrier densities (holes) in the range 2-8$\times10^{12}$ cm$^{2}$.}}
\label{Hall}
\end{figure}

Hall measurements show that the Hall mobility of the graphene sheets used in our diodes is in the range 1400-2100 cm$^2$/Vs, and that the sheets are hole doped with carrier densities in the range 2-8$\times10^{12}$ cm$^{-2}$ (Fig.~\ref{Hall}). The presence of extrinsic residual doping in exfoliated graphene has been previously reported\cite{novoselov} and attributed to residual water vapor (\textit{p} type) or NH$_3$ (\textit{n}-type). In both cases annealing reduces the concentration of the dopants and forces $E_F^{gr}$ closer to the neutrality point. For our CVD prepared graphene, the presence of residual impurity doping can be attributed to a lowering of $E_F^{gr}$ due to hole doping of the graphene during the Fe(III)NO$_3$ etching-transfer process\cite{Su}.

\subsection{C. Current-voltage measurements}
\begin{figure}[t]
\includegraphics[angle=0,width=0.5\textwidth]{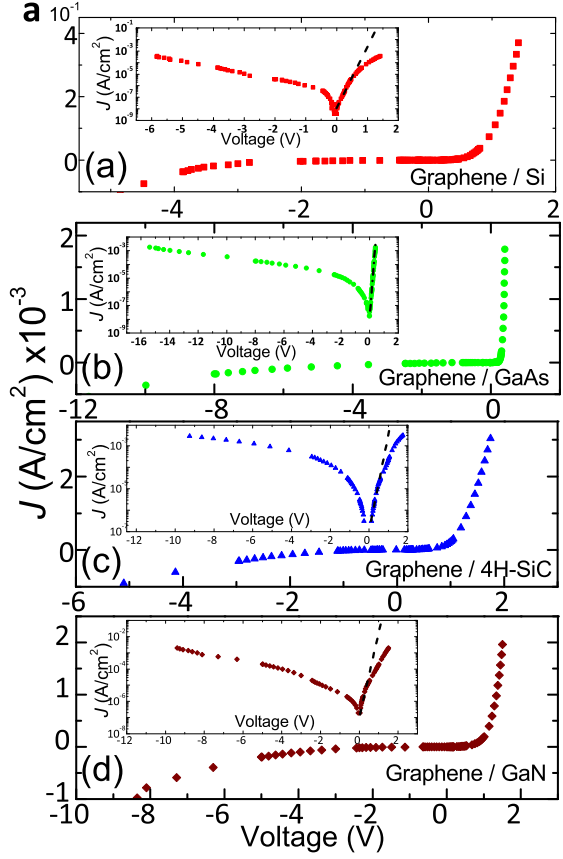}
\caption{{\footnotesize Room temperature current density-voltage characteristics show Schottky rectification at the (a) graphene/$n$-Si, (b) $n$-GaAs, (c) $n$-4H-SiC and (d) $n$-GaN interfaces. Insets: Semilogarithmic leaf plots, log$J$-$V$, reveal a thermionic emission dominated current density in forward bias that spans at least two decades of linearity (dotted lines) allowing us to extract out the Schottky barrier height recorded in Table.~I.}}
\label{IV}
\end{figure}

\begin{figure}[t]
\includegraphics[angle=0,width=0.5\textwidth]{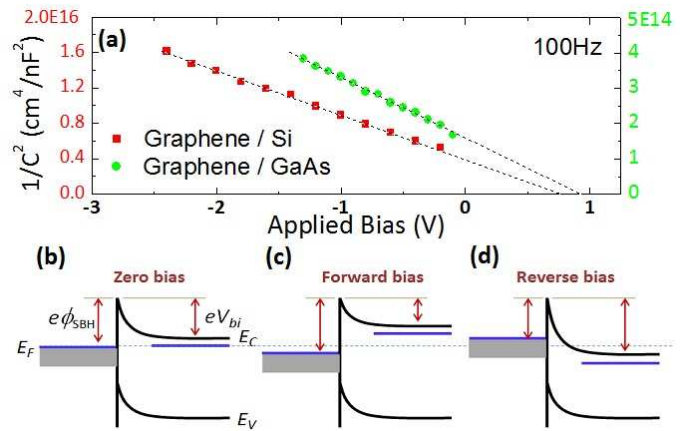}
\caption{{\footnotesize Capacitive response of the graphene based Schottky diodes, determination of the built-in potential, $V_{bi}$, and the proposed Schottky band diagram. (a) Plots of the inverse square capacitance ($1/C^2$) versus applied bias ($V$) graphene/$n$-Si (Red squares) and $n$-GaAs (green circles) at 300~K and 100~Hz show a linear dependence implying that the Schottky-Mott model provides a good description. The intercept on the abscissa gives the built-in potential ($V_{bi}$) which can be correlated to the Schottky barrier height while the slope of the linear fit gives $2/eN_D\epsilon_s \epsilon_0$. Extracted $\phi_{SBH}$ and $N_D$ values are listed in Table.~I}}
\label{CV}
\end{figure}

Schottky diodes are expected to pass current in the forward bias (semiconductor is negatively biased) while becoming highly resistive in reverse bias (semiconductor is positively biased). As seen in Fig.~\ref{IV}(a-d), $J$-$V$ (main panels) and log$J$-$V$ (insets) data taken on various graphene/\textit{n}-type semiconductor junctions display strong rectification. This rectification is a consequence of Schottky barrier formation at the interface when electrons flow from the semiconductor to the graphene as the Fermi energies equilibrate (Fig.~\ref{CV}(b)). 

In principle, any semiconductor with electron affinity ($\chi_{e}$) smaller than the work function of the metal $(\Phi_{metal})$ can create rectification at a metal-semiconductor (M-S) interface with Schottky barrier height, $\phi_{SBH} = \Phi_{metal}-\chi_{e}$, given by the Schottky-Mott model. Electron transport over the Schottky barrier at the M-S interface is well described by thermionic emission theory (TE) with the expression;
\begin{equation}  \label{richard}
J(T,V) = J_{s}(T) [\exp ({eV}/{\eta k_B T})-1]   ,
\end{equation}
where $J(T,V)$ is the current density across the graphene/semiconductor interface, \textit{V} the applied voltage, \textit{T} the temperature and $\eta$ the ideality factor \cite{sze}. The prefactor, $J_s (T)$ is the saturation current density and is expressed as $J_{s} = A^* T^2 \exp ({-e \phi_{SBH}}/{k_B T})$, where $e\phi_{SBH}$ is the zero bias Schottky barrier height (SBH) and $A^{*}$ is the Richardson constant.

\begin{figure}[t]
\includegraphics[angle=0,width=0.5\textwidth]{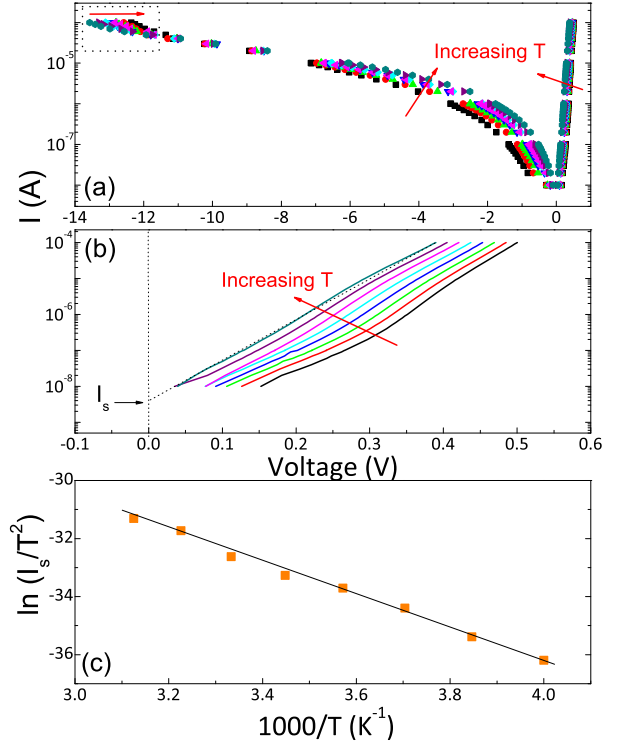}
\caption{{\footnotesize (a) The temperature dependence of the current ($I$) versus voltage ($V$) curves measured across a graphene/GaAs junction from 250~K up to 320~K with 10~K intervals separating each isotherm. 
The arrows indicate the direction of increasing temperature. (b) The temperature dependence of $I$-$V$ curves taken on graphene/GaAs junctions at different temperatures. (c) Extracted $I_s$ values from Fig.\ref{IV2}(b) are plotted in terms of ln$I_s$/$T^2$ versus 1000/T.}}
\label{IV2}
\end{figure}

When electronic transport across the barrier is dominated by thermionic emission described by Eq.~1, semilogarithmic plots of the $J$-$V$ curves should display a linear region in forward bias. As seen in the insets of Fig.~\ref{IV}(a-d) the overlying straight line segments of our measurements typically reveal 2-4 decades of linearity, thus allowing us to extract $J_s$ and $\eta$ for each diode. The deviations from linearity at higher bias are due to series resistance contributions from the respective semiconductors. The temperature-dependent data for the graphene/GaAs diode (Fig.~\ref{IV2}(a-b)) show that for both bias directions, a larger (smaller) current flows as the temperature is increased (decreased) and the probability of conduction electrons overcoming the barrier increases (decreases). In forward bias, the TE process manifests itself as linear \lq\lq log$J$-V curves\rq\rq (Fig.~\ref{IV2}(b)) and linear \lq\lq$\ln(I_{s}(T)/T^2)$ versus $T^{-1}$ curves\rq\rq (Fig.~\ref{IV2}(c)) where $I_s(T) = J_s(T)A$. The SBH is calculated directly from the slope of this linear dependence. By repeating these temperature-dependent measurements for the four different diodes, we find that the SBH ($\phi^{JV}_{SBH}$) values at the graphene/semiconductor interfaces are 0.86\,eV, 0.79\,eV, 0.91\,eV and 0.73\,eV for Si, GaAs, SiC and GaN respectively (Table~I). While the overall reverse current density increases as $T$ is increased, we notice that at high reverse bias the magnitude of the breakdown voltage $V_b$ decreases linearly with temperature (see boxed region in upper left hand corner of Fig.~\ref{IV2}(b)) implying that $V_b$ has a positive breakdown coefficient and that the junction breakdown mechanism is mainly avalanche multiplication\cite{sze}.

 \begin{table} 
 \caption{Extracted SBHs, doping densities, and corresponding graphene work function values on various graphene/semiconductor junctions}
 \centering
 \begin{tabular} {p{2.6cm} p{0.7cm} p{0.7cm} p{1.5cm} p{1.5cm} p{0.6cm}}
 \hline\hline
  & $\phi_{SBH}^{JV}$ & $\phi_{SBH}^{CV}$ & $N_{D}^{CV}$ & $N_{D}^{Hall}$ & $\Phi_{gr}$ \\ [1.0 ex] 
  \footnotesize Junction type & [eV] & [eV] & [cm$^{-3}$] & [cm$^{-3}$] & [eV] \\ [1.0 ex]
 \hline
 \footnotesize Graphene/\textit{n}Si & 0.86 & 0.92 & $4.0\times10^{15}$ & $3.0\times 10^{15}$ & 4.91 \\
 \footnotesize Graphene/\textit{n}GaAs & 0.79 & 0.91 & $3.5\times10^{16}$ & $3.0\times10^{16}$ & 4.89 \\
 \footnotesize Graphene/\textit{n}4H-SiC & 0.91 & \footnotesize ~N A & \footnotesize ~~~~~N A  & $1.0\times10^{16}$ & 4.31 \\
 \footnotesize Graphene/\textit{n}GaN & 0.73 & \footnotesize ~N A & \footnotesize ~~~~~N A & $1.0\times10^{17}$ & 4.83 \\
 \hline
 \end{tabular}
 \label{table}
 \end{table}
 
Schottky barrier values are well described using either the Bardeen or Schottky limits. In the Bardeen limit, the interface physics is mostly governed by interface states which, by accumulating free charge, change the charge distribution at the interface and cause $E_F$ of the semiconductor to be fixed (Fermi level pinning). Accordingly, the SBH shows weak dependence on the work function of the metals used for contacts, as is found for example in GaAs~\cite{sze}. On the other hand, the wide band gap semiconductors SiC and GaN are well described by the Schottky-Mott (S-M) limit,
\begin{equation}  \label{SM}
\phi_{SBH} = \Phi_{gr} - \chi_e   ,
\end{equation}
where $\Phi_{gr}$ is the work function of the graphene and $\chi$ is the electron affinity of the semiconductor. Using the extracted values of $\phi_{SBH}$, and electron affinity values ($\chi_{Si}\sim 4.05$~eV, $\chi_{GaAs}\sim 4.1$~eV, $\chi_{4H-SiC}\sim 3.4$~eV and $\chi_{GaN}\sim 4.1$~eV), we calculate $\Phi_{gr}$ (Table I). The calculated values of the work function are typically higher than the accepted values ($\sim$4.6~eV) of graphene when $E_F$ is at the Dirac point (K point). The deviation from this ideal graphene work function can be attributed to the lowering of $E_F$ due to hole doping of the graphene during the Fe(III)NO$_3$ etching-transfer process\cite{Su} (Fig.~\ref{Raman}(c-d)) together with the fact that the graphene is in physical contact with the gold electrodes\cite{Kim} (Fig.~\ref{Hall}).
 
Although the SBHs on Si, GaAs and GaN can be roughly explained within the S-M model, in reality GaAs surfaces have a high density of surface states and thus exhibit characteristic Fermi level pinning. In the Bardeen limit, GaAs based diodes generally have SBHs in the range of 0.75-0.85~eV as observed in our measurements, and proper interpretation of the SBH on GaAs/graphene junctions requires the Bardeen model. Subsequent to the placement of the graphene on the semiconductor surface, there is charge separation and concomitant formation of induced dipoles at the interface. According to bond polarization theory\cite{tung,tungrev}, the SBH is determined by charge separation at the boundary between the outermost layers of the metal (here, a single layer carbon sheet) and the semiconductor. Our results are in good agreement with the findings of our earlier work on graphite and MLG junctions where the layer in closest proximity to the semiconductor surface is a single sheet of carbon atoms\cite{tongay1,tongay2}. On the other hand, barriers formed on the 4H-SiC substrates give an unphysically low value for $\Phi_{gr}$ (see Table I) and therefore cannot be explained by either model. 

Next we turn our attention to reverse bias characteristics when the semiconductor (graphene) is positively (negatively) charged. In conventional metal-semiconductor Schottky diodes, the work function of the metal is pinned independent of bias voltage due to the high density of states at $E_F$ while in the reverse (forward) bias the Fermi energy of the semiconductor shifts down (up) allowing observed rectification via an increase (decrease) in the built-in potential ($V_{bi}$). Unlike conventional metals, graphene's work function ($\Phi_{gr}$) is a function of bias \cite{Sood}, and for large voltage values the SBH does not stay constant. When Schottky diodes are forward biased, they pass large currents at voltages well below $\sim$1~V and and small decreases in the Fermi level of graphene cannot be distinguished from voltage drops associated with a series resistance. Said in another way, the deviation from linearity in the semilogarithmic plots of Fig.~\ref{IV}(a-d) for forward bias could be due to a combination of a series resistance becoming important at high currents together with a small increase in $\Phi_{gr}$ and a downward shift in $E_F$ for the positively charged graphene. However, in reverse bias where the applied voltage can be larger than 10~V, $E_F$ starts changing dramatically\cite{Kim} and the fixed SBH assumption clearly no longer holds. In reverse bias when the graphene electrodes are negatively charged, $E_F$ increases and $\Phi_{gr}$ decreases causing the SBH height to decrease as the reverse bias is increased. As observed in the insets of Fig.~\ref{IV}(a-d) this effect causes the total reverse current to increase as the magnitude of the bias is increased, thus preventing the Schottky diode from reaching reverse current saturation. This non-saturating reverse current has not been observed in graphite based Schottky junctions due to the fixed Fermi level of graphite~\cite{tongay1}.

\subsection{D. Capacitance-voltage measurements}
Capacitance-voltage $C$-$V$ measurements made in the reverse bias mode are complementary to $J$-$V$ measurements and provide useful information about the distribution and density $N_D$ of ionized donors in the semiconductor and the magnitude of the built-in potential $V_{bi}$.
For a uniform distribution of ionized donors within the depletion width of the semiconductor, the Schottky-Mott relationship between $1/C^2$ and the reverse bias voltage $V_R$, satisfies the linear relationship, $1/C^2=2(V_R+V_{bi})/eN_D\epsilon_{s}\epsilon_0$, which as shown in Fig.~\ref{CV}(a) is observed to hold for graphene/GaAs and graphene/Si junctions. Linear extrapolation to the intercept with the abscissa gives the built-in potential, $V_{bi}$, which is related to $\phi_{SBH}$ via the expression, $\phi_{SBH}=V_{bi}+e^{-1}k_bT$ln$(N_{c}/N_{D})$~\cite{sze}. Here $N_c$ is the effective density of states in the conduction band, $N_D$ is the doping level of the semiconductor, and the slope of the linear fitting to $1/C^2$ versus $V_R$ gives the doping density of the semiconductor. We list $\phi_{SBH}^{CV}$ and $N_D$ values for the graphene/GaAs and graphene/Si junctions in Table.~I.  

We note from Table 1 that the extracted $\phi^{CV}_{SBH}$ values on the Si and GaAs junctions are generally higher than $\phi^{JV}_{SBH}$. The discrepancy between the SBHs determined by the two methods can be attributed to: (a) the existence of a thin oxide or residue at the graphene/semiconductor interface, and/or (b) Schottky barrier inhomogeneity. Graphene sheets transferred onto SiO$_2$ are known to have charge puddles mostly due to the inhomonegous doping either originating from natural graphite (mechanical exfoliation transfer) or from chemicals used during the graphene production or transfer (CVD graphene transfer) process. Since the SBH is sensitive to the E$_F$ of graphene, patches with different charge densities (doping) are expected to have an impact on the SBH and hence the $J$-$V$ characteristics of the graphene diodes. 

An important difference between the $C$-$V$ and $J$-$V$ techniques is that the $C$-$V$ measurements probe the average junction capacitance at the interface thereby yielding an average value for the SBH, while the $J$-$V$ measurements give a minimum value for the SBH, since electrons with thermionic emission probabilities exponentially sensitive to barrier heights choose low barrier patches (less \textit{p}-doped graphene patches) over higher patches (more p-doped graphene patches)\cite{tungrev}. While $C$-$V$ measurements give reasonable values of the SBH for graphene/GaAs and graphene/Si, we have not been able to obtain reliable $C$-$V$ measurements for graphene deposited on GaN and SiC because of high series resistance in these wide band gap semiconductors. 

The linearity of the $C$-$V$ measurements shown in Fig.~\ref{CV} is consistent with the Schottky-Mott model and the abrupt junction approximation, which assumes that the density of ionized donors $N_D$ is constant throughout the depletion width of the semiconductor. This good agreement invites a more quantitative analysis of the Fermi energy shifts in the graphene that are the source of the non-saturating reverse bias currents discussed in the previous subsection. We begin by writing the electron charge density per unit area $Q$ on the graphene as 
\begin{equation}  \label{QCV}
Q=en_{induced}=C_{dep}(V_{bi}+V_{R}) ,
\end{equation}
where 
\begin{equation}  \label{Cdep}
C_{dep}=\sqrt{\frac{e\epsilon_s\epsilon_0N_D}{2\left(V_{bi}+V_R\right)}},
\end{equation}
is the Schottky-Mott depletion capacitance, $n_{induced}$ is the number of electrons per unit area and $V_R$ is the magnitude of the reverse bias voltage. Combining these two equations gives the result, 
\begin{equation}  \label{ninduced}
n_{induced}=\sqrt{\epsilon_s \epsilon_0 N_D(V_{bi}+V_R)/{2e}}   .
\end{equation}

The above expression provides an estimate of the number of carriers per unit area associated with the electric field within the depletion width but does not account for extrinsic residual doping described by the carrier density $n_0$ on the graphene before making contact with the semiconductor. The processing steps used to transfer the CVD grown graphene from Cu substrates to semiconductor surfaces typically results in \textit{p}-doped material with $n_0 \sim 5 \times 10^{12}$cm$^{-2}$ as inferred from Hall data (Fig.~\ref{Hall}) taken at 300~K. Accordingly, the final carrier density including contributions from the as-made graphene and the charge transfers associated with the Schottky barrier ($V_{bi}$ and the applied voltage $V_R$) reads as,
\begin{equation}  \label{nafter}
n_{final}=n_{0}-n_{induced},
\end{equation}
Using the well-known expression for graphene's Fermi energy\cite{castroneto} we can write
\begin{equation}  \label{EF}
E_F=-\hbar\left|v_F\right|k_F=-\hbar\left|v_F\right|\sqrt{\pi(n_{0}-n_{induced})},
\end{equation}
which in combination with Eq.~\ref{ninduced} becomes
\begin{equation}  \label{EFfinal}
E_F=-\hbar\left|v_F\right|\sqrt{\pi (n_0 - \sqrt{\epsilon_s \epsilon_0 N_D(V_{bi}+V_R)/{2e}})},
\end{equation}

To calculate typical shifts in $E_F$, we use parameter values $\epsilon_0 = 8.84 \times 10^{-14}$~F/cm$^2$, $\hbar = 6.5 \times 10^{-16}$~eV\,s, $e = 1.6 \times 10^{-19}$~C, $v_F = 1.1 \times 10^8$~cm/s, $V_{bi} \sim 0.6$~V and $\epsilon_s \sim 10$ for a typical semiconductor. Thus the Fermi energy of the as-made graphene with $n_0 \sim 5 \times 10^{12}$~cm$^{-2}$ is calculated from $E_F=-\hbar\left|v_F\right|\sqrt{\pi n_0}$ to be $-0.287$~eV below the charge neutrality point, a shift associated with the aforementioned \textit{p}-doping during processing. When the graphene is transferred to the semiconductor, equilibration of the chemical potentials and concomitant formation of a Schottky barrier (Fig.~\ref{CV}) results in a transfer of negative charge to the graphene and an increase in $E_F$ (calculated from Eq.~\ref{EFfinal} for $V_R = 0$) to be in the range 3 to 11~meV for $N_D$ in the range $1\times10^{16}$ to $1\times 10^{17}$~cm$^{-3}$. The application of a typical 10~V reverse bias (see Figs.~\ref{IV} and ~\ref{IV2}) creates significantly larger Fermi energy shifts which from Eq.~\ref{EFfinal} give $E_F$ in the range $-0.271$ to $-0.233$~eV for the same factor of ten variation in $N_D$. The corresponding shifts from the pristine value of $-0.287$~eV are in the range 15~-~53~meV and thus bring E$_F$ closer to the neutrality point. These numerical calculations show that for our \textit{n}-doped semiconductors, it is relatively easy to induce Fermi energy shifts on the order of 50~meV with the application of a sufficiently high reverse bias voltage. An upward shift in $E_F$ of 50~meV causes a reduction in $\Phi_{gr}$ by the same amount. Since the electron affinity of the semiconductor remains unchanged, the Schottky-Mott constraint of Eq.~\ref{SM} enforces the same reduction in $\phi_{SBH}$ thus leading to a greater than 5\% reduction in the measured SBH's shown in Table I. We note that the induced shift in graphene's E$_F$ as determined by the in-situ Raman spectroscopy measurements (Fig.\ref{RamanBias}) is larger ($\Delta$E$_F$$\sim$200-500$\thinspace$meV) than our theoretical estimation ($\Delta$E$_F$$\sim$50$\thinspace$meV).

The discrepancy between the theoretical estimate of $\Delta$E$_F$ and the experimental values might be attributed to: (1) the existence of an interface capacitance induced by dipoles at the graphene/semiconductor interface (within bond polarization theory) causing deviation from the ideal Schottky-Mott capacitance relation given by Eq.~\ref{Cdep} and (2) the estimate of $\Delta$E$_F$ using relative peak shifts in the G and 2D peak positions for graphene deposited on Si/SiO$_2$~\cite{Sood} might be different than the change in Fermi level for graphene transferred onto semiconductors.

\subsection{E. Modification of thermionic emission theory}

As discussed in the previous sections, since the E$_F$ of graphene electrode is sensitive to the applied bias across the graphene/semiconductor interface, the SBH at the interface becomes bias dependent especially for large reverse voltages. However, extracting the SBH from $J$-$V$ characteristics using Eq.~\ref{richard} which involves extrapolating current density to zero bias saturation current (J$_s$) yields the putative zero bias barrier height (Table.1). In this section, we present a simple modification to the Richardson equation (Eq.~\ref{richard}) considering the shift in E$_F$ of graphene induced by applied bias. The modified Richardson equation preserves the original functional form of Eq.~\ref{richard} but allows one to estimate the SBH at fixed voltages.

The voltage-dependent SBH ($\Phi_{SBH}(V)$) can be written as,
\begin{equation}
\label{deltaSBH1}
e \Phi_{SBH} = e \Phi^0_{SBH}+e\Delta\Phi_{SBH}\left(V\right)\\
= e \Phi^0_{SBH}-\Delta E_{F}\left(V\right)\\
\end{equation}
where e$\Phi$$^0_{SBH}$ is the zero bias SBH and e$\Delta\Phi_{SBH}(V)$ is the correction to the SBH at fixed voltage V.
The change in the Fermi energy $\Delta E_{F}\left(V\right)$ is opposite to e$\Delta\Phi_{SBH}(V)$, i.e., $\Delta E_{F}\left(V\right) = -e\Delta \Phi_{SBH}(V)$, as seen in Fig.~\ref{CV}b. 
Thus for reverse bias (addition of electrons to the graphene) we use Eq.~\ref{ninduced} in Eq.~\ref{EF}, 
together with the inequality $n_{induced}<< n_0$ to calculate 
\begin{equation}
\label{deltaSBH}
\begin{split}
& e\Delta\Phi_{SBH}(V_R) = - \Delta E_F(V_R) = \\
& \hbar v_F \left[ \sqrt{\pi(n_0 - n_{induced})} -\sqrt{\pi n_0}\right]\\
& \approx -\frac{1}{2} \hbar v_F \sqrt{\pi n_0} \frac{n_{induced}}{n_0}\\
& = -\frac{1}{2} \hbar v_F \sqrt { \frac {\pi \epsilon_s \epsilon_0 N_D (V_{bi}+V_R)}{2en_0}}
\end{split}
\end{equation}

Adding the reverse and forward current densities as is done in standard treatments of the diode equation\cite{sze} yields the total current density across the graphene/semiconductor interface,
\begin{equation}
\label{FR1}
\begin{split}
& J(V)= \\
& A^* T^2 \texttt{exp}\left(-\frac{e\Phi^0_{SBH}+e\Delta\Phi_{SBH}\left(V\right)}{k_B T}\right) \biggl[\texttt{exp}\left(\frac{eV}{k_B T}\right)-1\biggr]
\end{split}
\end{equation}
Here, we note that the original form of the Richardson equation is preserved with slight modifications to the saturation current term which is given as;
\begin{equation}
\label{final2}
J_s=A^* T^2 \texttt{exp}\left(-\frac{e\Phi^0_{SBH}+e\Delta\Phi_{SBH}(V)}{k_B T}\right)
\end{equation}
with $\Delta\Phi_{SBH}(V)$ for reverse bias given by Eq.~\ref{deltaSBH}.

In our conventional $J$-$V$ analysis using Eq.~\ref{richard}, the zero-bias saturation current J$_s$ is extracted by extrapolating the current density to zero bias limit. In this limit, the correction to the SBH is expected to be zero, since the graphene is not subject to applied bias and hence the Fermi level does not shift from the original value. However, using the extrapolated zero-bias saturation current density, one can extract out the SBH and the correction to the SBH at fixed bias (V) can be taken into account by the additional term ($\Delta\Phi_{SBH}(V)$ in Eq.~\ref{final2}).

\section{Conclusion}
In summary, we have used current-voltage and capacitance-voltage measurements to characterize the Schottky barriers formed when graphene, a zero-gap semiconductor, is placed in intimate contact with the \textit{n}-type semiconductors: Si, GaAs, GaN and SiC. The good agreement with Schottky-Mott (S-M) physics within the context of bond-polarization theory is somewhat surprising since the S-M picture has been developed for metal-semiconductor interfaces, not for single atomic layer ZGS-semiconductor interfaces discussed here. Moreover, due to a low density of states, graphene's Fermi level shifts during the charge transfer across the graphene-semiconductor interface. This shift does not occur at metal-semiconductor or graphite-semiconductor interfaces where $E_F$ remains fixed during Schottky barrier formation and the concomitant creation of a built-in potential, $V_{bi}$ with associated band bending (see Fig.~\ref{CV}). Another major difference becomes apparent when under strong reverse bias. According to our in-situ Raman spectroscopy measurements, large voltages across the graphene/semiconductor interface change the charge density and hence the Fermi level of graphene as determined by relative changes in the G and 2D peak positions. The bias-induced shift in the Fermi energy (and hence the the work function) of the graphene causes significant changes in the diode current. Considering changes in the barrier height associated with bias induced Fermi level shift, we modify the thermionic emission theory allowing us to estimate the change in the barrier height at fixed applied bias. The rectification effects observed on a wide variety of semiconductors suggest a number of applications, such as to sensors where in forward bias there is exponential sensitivity to changes in the SBH due to the presence of absorbates on the graphene or to MESFET and HEMT devices for which Schottky barriers are integral components. Graphene is particularly advantageous in such applications because of its mechanical stability, its resistance to diffusion, its robustness at high temperatures and its demonstrated capability to embrace multiple functionalities.

We thank N. Savage for her help during preparation of the diagrams used in this manuscript and Prof. A. Rinzler and Dr. M. McCarthy for technical assistance. This work is supported by the Office of Naval Research (ONR) under Contract Number 00075094 (BA) and by the National Science Foundation (NSF) under Contract Number 1005301 (AFH).








\end{document}